\documentclass[aps,pre,amsmath,amsfonts,amssymb,twocolumn,superscriptaddress,showpacs]{revtex4-1}
\usepackage{amssymb,amsmath,amsfonts}
\usepackage{graphicx,psfrag,color}
\usepackage{dcolumn}
\usepackage{bbm}
\usepackage{mathbbol}
\usepackage{braket}
\usepackage{verbatim}

\newcommand*{\balancecolsandclearpage}{%
  \close@column@grid
  \clearpage
  \twocolumngrid
}

\begin{document}
\flushbottom

\title{Stiffness in Confined Quantum Walks} 

\author{Pablo G. M. Dantas}
\affiliation{Departamento de F\'isica, Universidade do Estado de Santa Catarina, 89219-710, Joinville, SC, Brazil}

\author{Rafael Vieira}
\affiliation{Departamento de F\'isica, Universidade do Estado de Santa Catarina, 89219-710, Joinville, SC, Brazil}

\author{Edgard P. M. Amorim}
\email{edgard.amorim@udesc.br} 
\affiliation{Departamento de F\'isica, Universidade do Estado de Santa Catarina, 89219-710, Joinville, SC, Brazil}

\date{\today}

\begin{abstract}
We investigate boundary-induced stiffness in discrete-time quantum walks confined to a finite lattice by unitary reflecting boundaries implemented through coin-flip operations. We define a dimensionless spreading stiffness from the instantaneous variation of the spatial width relative to the asymptotic spreading rate of the corresponding unconfined Hadamard walk and compare localized, Gaussian, and truncated-uniform initial states. The long-time response retains a strong dependence on the initial spatial profile: the localized and truncated-uniform states show an overall increase in stiffness with increasing confinement size, whereas the Gaussian state exhibits a nonmonotonic dependence with a broad maximum at intermediate sizes. The first and second stiffness moments satisfy the expected asymptotic consistency relation in the high-stiffness regime. The normalized Kullback--Leibler divergence reveals a slower spatial reorganization and qualitatively different finite-size trends, showing that dynamical stiffness and spatial homogenization are distinct properties. These results indicate that unitary confinement produces an intrinsically nonstationary regime whose long-time properties emerge from the interplay between boundary reflections, system size, and the spatial structure of the initial state.
\end{abstract}


\maketitle

\section{Introduction} \label{sec:intro}

Random walks~\cite{Pearson1905problem} constitute a fundamental framework for describing stochastic transport processes and have found widespread applications across many areas of science~\cite{Masuda2017random}. They provide a simple yet powerful model for diffusion and transport phenomena~\cite{Weiss1994aspects,Hughes1995random}, including transport in disordered media~\cite{Bouchaud1990anomalous}. In their simplest form, the probabilistic nature of classical random walks leads to diffusive behavior, characterized by a linear growth of the position variance in time.

When random walks are confined by reflecting boundaries, their dynamics undergoes a qualitative change due to the interplay between diffusion and confinement. For an unconfined one-dimensional random walk, occupation-time statistics constitute an important class of Brownian functionals~\cite{Majumdar2005brownian} and are governed by the L\'evy--arcsine law~\cite{Levy1940sur,Morters2010brownian}. Reflecting boundaries suppress this behavior and drive a crossover from the characteristic bimodal distribution to a unimodal distribution centered around equal occupation~\cite{Kaldasch2022stiffness}. In this classical setting, the notion of random-walk stiffness is associated with occupation-time statistics, providing the motivation for the quantum formulation considered below.

Quantum walks provide the quantum counterpart of classical random walks and can be formulated in both discrete-time and continuous-time frameworks~\cite{Aharonov1993quantum,Farhi1998quantum}. In the discrete-time formulation, the evolution is governed by unitary coin and conditional-shift operations, with interference playing a central role. In contrast to classical walks, quantum walks exhibit ballistic spreading, with the position variance growing quadratically in time~\cite{Ambainis2001one,Kempe2003quantum,VenegasAndraca2012quantum}. Quantum walks have provided a framework for quantum search algorithms~\cite{Shenvi2003quantum,Tulsi2008faster,Portugal2013quantum} and universal quantum computation~\cite{Childs2009universal,Lovett2010universal}, and have also been employed as models of coherent transport~\cite{Mulken2011continuous}. Notably, quantum walks have also been experimentally implemented across a variety of physical platforms~\cite{Wang2014physical,Flamini2019photonic}.

From a dynamical perspective, quantum walks are highly sensitive to the initial state~\cite{Tregenna2003controlling,Grimmett2004weak} and can be strongly modified by different forms of disorder introduced through their evolution operators. Dynamic disorder can enhance coin-position entanglement and even drive it toward its maximal asymptotic value~\cite{Vieira2013dynamically,Orthey2019weak}, while comparisons between dynamic, fluctuating, and static disorder reveal markedly different entanglement and transport properties~\cite{Vieira2014entangling}. Spatially correlated disorder can induce localization-delocalization transitions and intermediate transport regimes~\cite{Mendes2019localization}, while aperiodic spatial inhomogeneities can modify transport and enhance coin-position entanglement~\cite{Buarque2019aperiodic}. Remarkably, even a single local defect can induce localization, trapping, and substantial changes in the spreading dynamics~\cite{Wojcik2012trapping,Li2013position,Zhang2014one}. These features highlight the sensitivity of quantum walks to local perturbations and make them versatile systems for exploring nonequilibrium quantum dynamics.

Reflecting boundaries in quantum walks require a unitary prescription, since probability must be conserved throughout the evolution. Local Pauli-$X$ operations have previously been employed to reflect, confine, and control the propagation of quantum-walk wave packets~\cite{Ghizoni2019trojan,Vieira2021quantum,Engster2024high}. These operations interchange the left- and right-moving coin components whenever the walker reaches a boundary, thereby reversing the direction of propagation while preserving coherence. The resulting dynamics remains fully unitary, but translational invariance is broken, and the spectral properties of the finite evolution operator encode the effect of confinement. In this way, the boundary does not absorb or destroy probability; instead, it acts as a coherent reflector. Such a construction provides a simple unitary framework for investigating how repeated reflections modify interference, spreading, and the stiffness of the quantum walk.

Despite the extensive literature on quantum walks and previous studies of unitary confinement and controlled spreading, the notion of stiffness in confined quantum walks remains largely unexplored. In particular, the relation between reflecting boundaries, spreading rate, and the long-time redistribution of probability has not, to our knowledge, been systematically investigated in the quantum regime. In the standard quantum-mechanical framework, trajectory information is not available independently of measurement, and extracting it generally requires interventions that modify the unitary dynamics~\cite{Peres1993quantum}. Consequently, trajectory-based occupation-time statistics are not directly available for the coherent quantum walk considered here.

In this work, we therefore formulate stiffness operationally through the resistance of the spatial width to temporal variation under confinement. We introduce a dimensionless spreading stiffness by comparing its instantaneous variation with the asymptotic spreading scale of the corresponding unconfined Hadamard walk. We investigate the resulting stiffness for localized and delocalized initial states with distinct spatial profiles, allowing us to assess how the initial spatial structure influences the response to confinement. In this way, we address the gap between the classical notion of stiffness and coherent confined quantum dynamics, providing new insights into the interplay between confinement, spreading, and coherence in quantum transport processes.

This paper is organized as follows. In Sec.~\ref{secII}, we review the classical random walk with reflecting boundaries and its diffusive approach to a uniform stationary distribution, which provides the reference behavior for the quantum problem. In Sec.~\ref{secIII}, we introduce the discrete-time quantum-walk formalism, the implementation of reflecting boundaries through coin-flip operations, the dynamical observables, and the initial states considered in this work. In Sec.~\ref{secIV}, we introduce the spreading-stiffness and Kullback--Leibler diagnostics and present the numerical results, including the confinement dynamics and the finite-size behavior of the corresponding long-time statistics. Finally, in Sec.~\ref{secV}, we summarize our main conclusions. 

\section{Classical random walk with reflecting boundaries}\label{secII}

Before introducing the confined quantum walk, it is useful to recall the classical diffusion with reflecting boundary conditions. This provides the reference behavior against which the quantum dynamics will be compared. The derivation of the diffusion equation from the continuum limit of an unbiased random walk, as well as the solution of the corresponding boundary-value problem, is standard and can be found in classical treatments of diffusion~\cite{Crank1956mathematics}.

We consider a one-dimensional probability density \(p(x,t)\) confined to the interval \([-1,1]\). In the diffusion limit, the dynamics is governed by
\begin{equation}
\partial_t p(x,t)=D\,\partial_x^2 p(x,t),
\label{eq:diffusion}
\end{equation}
where \(D\) is the diffusion coefficient. Reflecting boundaries impose a vanishing probability current at the edges. Since \(J(x,t)=-D\,\partial_x p(x,t)\), this condition is equivalent to the Neumann boundary conditions $\partial_x p(\pm1,t)=0$. For an initially localized walker, $p(x,0)=\delta(x)$, the solution is even under \(x\to -x\). Therefore, only the even Neumann modes contribute, and the probability density can be written as
\begin{equation}
p(x,t)=\frac12+\sum_{n=1}^{\infty}\cos(n\pi x)\,
e^{-Dn^2\pi^2t}.
\label{eq:classical_solution}
\end{equation}
The zero mode gives the stationary distribution, while all nonuniform modes decay exponentially. Hence, $p(x,t\to\infty)\to 1/2$ showing that the classical confined process relaxes to the uniform distribution on \([-1,1]\). This behavior is the classical signature of confinement by reflecting boundaries: large excursions are suppressed, all nonuniform spatial modes decay, and only the stationary mode survives. 

\begin{figure}[htpb]
\centering
\includegraphics[width=0.45\textwidth]{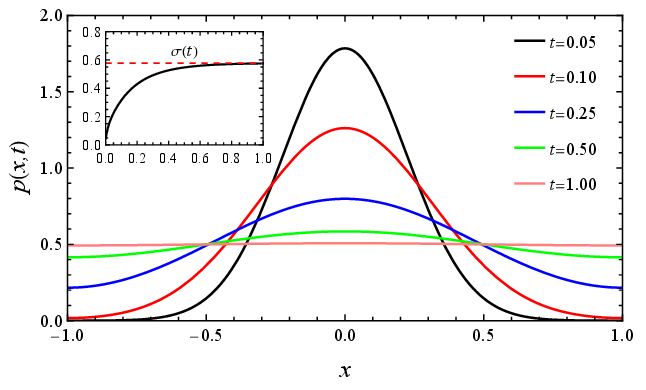}
\caption{Classical diffusion with reflecting boundaries on $[-1,1]$, with diffusion coefficient $D=1$. The main panel shows $p(x,t)$ from Eq.~\eqref{eq:classical_solution} at different times. The distribution approaches the uniform profile $p(x)=1/2$ as nonuniform modes decay. The inset shows the standard deviation $\sigma(t)$ (black line), which approaches the asymptotic value $\sigma_\infty=1/\sqrt{3}$ (red dashed line).}
\label{Fig1}
\end{figure}

The spreading of the classical probability density is characterized by the variance $\sigma^2$. For the initial condition considered here, since \(\langle x\rangle=0\), we have $\sigma^2=\langle x^2\rangle$, and using Eq.~\eqref{eq:classical_solution} one obtains
\begin{equation}
\sigma^2(t)=\frac{1}{3}+\frac{4}{\pi^2}\sum_{n=1}^{\infty}\frac{(-1)^n}{n^2}e^{-Dn^2\pi^2 t}.
\end{equation}
In the long-time limit, all decaying modes vanish and $\sigma(t\to\infty)\to 1/\sqrt{3}$, which is the standard deviation of a uniform distribution on \([-1,1]\). Figure~\ref{Fig1} illustrates the spreading of the classical probability distribution and the standard deviation over time.

In contrast to classical random walks, where stiffness can be characterized through the occupation-time distribution and its relation to the Lévy arcsine law~\cite{Levy1940sur,Morters2010brownian}, a coherent quantum walk does not possess well-defined trajectories. We therefore formulate stiffness at the level of the probability distribution generated by the unitary dynamics. In the quantum setting considered here, stiffness is formulated through the resistance of the spatial width to temporal variation under confinement, while spatial redistribution relative to a uniform reference is characterized separately through the Kullback--Leibler divergence.

\section{Methodology} \label{secIII}

In this section, we present the theoretical framework used to analyze the dynamics and stiffness of quantum walks in the presence of reflecting boundaries.

\subsection{Discrete-time quantum walks}

Quantum walks describe the coherent propagation of a quantum particle on a discrete lattice, where interference effects lead to transport properties markedly different from their classical counterparts. Mathematically, the state belongs to a bipartite Hilbert space $\mathcal{H} = \mathcal{H}_c \otimes \mathcal{H}_p$, where $\mathcal{H}_c$ denotes the coin space spanned by the two-dimensional basis $\{\ket{L},\ket{R}\}$ and $\mathcal{H}_p$ is the position space spanned by the basis $\{\ket{j}\}$ with $j\in \mathbb{Z}$. A general initial state can be written as
\begin{equation}
\ket{\Psi(0)} = \sum_{j=-\infty}^{\infty}\left[\psi_L(j,0)\,\ket{L}+\psi_R(j,0)\,\ket{R}\right]\otimes\ket{j},
\end{equation}
where $\psi_L(j,0)$ and $\psi_R(j,0)$ are complex amplitudes satisfying the normalization condition
\begin{equation}
\sum_j\left[|\psi_L(j,0)|^2+|\psi_R(j,0)|^2\right]=1.
\end{equation}

The time evolution of the discrete-time quantum walk is governed by the repeated application of a unitary operator $U$, such that
\begin{equation}
\ket{\Psi(t)}=U^t \ket{\Psi(0)}.
\end{equation}
For a spatially homogeneous coin, the one-step evolution operator is
\begin{equation}
U=S\left(C\otimes I_p\right),
\end{equation}
where $C$ acts on the coin space, $I_p$ is the identity operator on the position space, and $S$ is the conditional shift operator,
\begin{equation}
S\!=\! \sum_j\left(\ket{L}\bra{L} \otimes \ket{j-\!1}\bra{j}\!+\!\ket{R}\bra{R} \otimes \ket{j\!+1}\bra{j}\right).
\end{equation}
Thus, the walker moves one lattice site to the left (right) when the coin state is $\ket{L}$ ($\ket{R}$), with the conditional shift generally entangling the internal coin and external position degrees of freedom when the coin state is in a superposition.

\subsection{Reflecting boundaries via coin-flip operators}

We introduce reflecting boundary conditions by replacing the spatially homogeneous coin $C$ with a position-dependent quantum coin. Let $\mathcal{B}=\{-a,\ldots,a\}$ denote the bulk region, containing $N=2a+1$ sites, and let
$\mathcal{W}=\{-a-1,a+1\}$ denote two auxiliary reflecting sites. The local coin is defined as
\begin{equation}
C_j=
\begin{cases}
H, & j\notin\mathcal{W},\\[2mm]
X, & j\in\mathcal{W},
\end{cases}
\end{equation}
where
\begin{equation}
H=\frac{1}{\sqrt{2}}
\begin{pmatrix}
1&1\\
1&-1
\end{pmatrix},
\qquad
X=\begin{pmatrix}
0&1\\
1&0
\end{pmatrix}
\end{equation}
are the Hadamard and Pauli-$X$ operators, respectively, with the latter acting as a coin flip (equivalently, a NOT gate in the $\{\ket{L},\ket{R}\}$ basis). The corresponding position-dependent coin operator can then be written as
\begin{equation}
C_{\mathrm{p}}=\sum_{j\in\mathbb{Z}} C_j\otimes\ket{j}\bra{j}.
\end{equation}
The one-step evolution operator then becomes
\begin{equation}
U=SC_{\mathrm{p}},
\end{equation}
and since both $S$ and $C_{\mathrm{p}}$ are unitary, $U$ is also unitary on $\mathcal{H}_c\otimes\mathcal{H}_p$.

The sites in $\mathcal{W}$ act as coherent reflectors and should therefore be distinguished from the $N$ bulk sites. Starting from an initial state supported on $\mathcal{B}$, the only dynamically accessible components at the auxiliary sites are the outward-moving states $\ket{L,-a-1}$ and $\ket{R,a+1}$. The coin-flip operation reverses their direction of propagation, such that
\begin{equation}
U\ket{L,-a-1}=\ket{R,-a},
\qquad
U\ket{R,a+1}=\ket{L,a}.
\end{equation}
Consequently, an amplitude reaching either auxiliary site is coherently returned to the bulk on the following time step, and no amplitude generated from an initial state supported on $\mathcal{B}$ propagates beyond $j=\pm(a+1)$. The dynamically generated subspace is therefore invariant under the evolution, providing a unitary realization of confinement. Throughout this work, $N=2a+1$ denotes the number of bulk sites, while the two auxiliary sites implement the reflecting boundaries. The resulting confinement preserves coherence while breaking translational invariance and thereby modifies the interference structure of the walk.

\subsection{Dynamical observables}

At an arbitrary time $t$, the normalized state of the walk can be written as
\begin{equation}
\ket{\Psi(t)}=\sum_{j}\left[\psi_L(j,t)\ket{L}+\psi_R(j,t)\ket{R}\right]
\otimes\ket{j},
\end{equation}
where $\psi_L(j,t)$ and $\psi_R(j,t)$ are the probability amplitudes associated with the left and right coin states, respectively. The spatial probability distribution is obtained by tracing out the coin degree of freedom,
\begin{equation}
P(j,t)=|\psi_L(j,t)|^2+|\psi_R(j,t)|^2.
\end{equation}
For the spreading analysis, the spatial moments are evaluated over the complete dynamically accessible region, including the two auxiliary reflecting sites,
\begin{equation}
\langle j^n\rangle_t=\sum_{j\in\mathcal{B}\cup\mathcal{W}}j^n P(j,t).
\end{equation}
The spatial standard deviation, which we use as a measure of the distribution width, is then defined as
\begin{equation}
\sigma(t)=\sqrt{\langle j^2\rangle_t-\langle j\rangle_t^2}.
\end{equation}
Thus, $\sigma(t)$ characterizes the full unitary spatial dynamics over the dynamically accessible region, including any transient occupation of the auxiliary reflecting sites.

The internal coin dynamics can also be characterized by tracing out the position degree of freedom. The reduced density matrix of the coin is therefore
\begin{equation}
\rho_c(t)=\mathrm{Tr}_{j}\left[\ket{\Psi(t)}\bra{\Psi(t)}\right]=
\begin{pmatrix}
\rho_{LL}(t) & \rho_{LR}(t)\\
\rho_{RL}(t) & \rho_{RR}(t)
\end{pmatrix},
\end{equation}
with matrix elements
\begin{align}
\rho_{LL}(t)&=\sum_j\left|\psi_L(j,t)\right|^2,\\
\rho_{RR}(t)&=\sum_j\left|\psi_R(j,t)\right|^2,\\
\rho_{LR}(t)&=\sum_j\psi_L(j,t)\psi_R^*(j,t),\\
\rho_{RL}(t)&=\rho_{LR}^*(t),
\end{align}
where the diagonal elements $\rho_{LL}$ and $\rho_{RR}$ represent the populations of the two internal coin states. The departure from an equally populated coin state can therefore be quantified by the population-imbalance contribution
\begin{equation}
\mathcal{P}(t)=\left[\rho_{LL}(t)-\frac{1}{2}\right]^2\!=\frac{1}{4}\left[\rho_{LL}(t)-\rho_{RR}(t)\right]^2.
\end{equation}
Thus, $\mathcal{P}(t)=0$ corresponds to equal populations of the left and right coin states, whereas larger values indicate an increasing population imbalance.

The off-diagonal elements quantify the coherence between the two coin states. In the present analysis, the coherence contribution is defined as
\begin{equation}
\mathcal{C}(t)=\left|\rho_{LR}(t)\right|^2=\left|\sum_j\psi_L(j,t)\psi_R^*(j,t)\right|^2.
\end{equation}
The positivity of the reduced density matrix requires $|\rho_{LR}(t)|^2\leq\rho_{LL}(t)\rho_{RR}(t)$ and since $\rho_{LL}(t)+\rho_{RR}(t)=1$, the coherence contribution is bounded by $0\leq \mathcal{C}(t)\leq \frac{1}{4}$.

The population and coherence contributions enter naturally in the eigenvalues of the reduced coin density matrix, and
its two eigenvalues can be written as
\begin{equation}
\lambda_{\pm}(t)=\frac{1}{2}\pm\sqrt{\mathcal{P}(t)+\mathcal{C}(t)}.
\end{equation}
This decomposition makes it possible to distinguish the contributions arising from population imbalance and coin-state coherence to the reduced-state spectrum and, consequently, the coin-position entanglement is given by the von Neumann entropy
\begin{equation}
S_E(t)=-\lambda_+(t)\log_2\lambda_+(t)-\lambda_-(t)\log_2\lambda_-(t).
\end{equation}
The entanglement entropy satisfies $0\leq S_E(t)\leq1$, with $S_E=0$ corresponding to a separable coin-position state and $S_E=1$ to maximal entanglement between coin and position, for which $\lambda_{\pm}=1/2$. 

\subsection{Initial states}

The quantum-walk initial state is a coin state placed over one or more discrete lattice positions. A general coin state has the form
\begin{equation}
\ket{\psi}_{c} = \cos\left(\frac{\theta}{2}\right)\ket{L}
+ e^{i\varphi}\sin\left(\frac{\theta}{2}\right)\ket{R},
\end{equation}
where $\theta \in [0,\pi]$ and $\varphi \in [0,2\pi]$ are respectively the polar and azimuthal angles in the Bloch-sphere representation~\cite{Nielsen2010quantum}. 

The choice of the initial coin state has a direct influence on the subsequent dynamics~\cite{Tregenna2003controlling,Grimmett2004weak}. For the Hadamard walk, the coin states defined by $(\theta,\varphi)=(0,0)$ and $(\pi,0)$ generate probability distributions biased in opposite directions. In contrast, the choice $(\theta,\varphi)=(\pi/2,\pi/2)$ gives the specific state,
\begin{equation}
\ket{\chi} =\frac{1}{\sqrt{2}}(\ket{L}+i\ket{R}).
\end{equation}
which removes the directional bias induced by the coin and yields a symmetric position distribution when the walker is initially centered at the origin and the boundaries are placed symmetrically. For the sake of simplicity, we adopt this symmetric coin state throughout this work.

The first initial state is localized at the origin given by
\begin{equation}
\ket{\Psi(0)}_L =\ket{\chi}\otimes\ket{0}.
\end{equation}
Moreover, it is instructive to consider delocalized states with a finite spatial width. Such extended initial states are experimentally accessible in quantum-walk platforms, where spatial superpositions and wave packets with controlled profiles have been realized~\cite{Cardano2015quantum,Su2019experimental,Derrico2020two}. A particularly relevant choice is a Gaussian wave packet, which allows one to control the initial spread of the walker and study its influence on the ensuing dynamics~\cite{Valcarcel2010tailoring,Orthey2017asymptotic,Orthey2019connecting}. For a bulk region containing $N=2a+1$ sites, with $j\in[-a,a]$, we also consider an initial state of the form
\begin{equation}
\ket{\Psi(0)}_G = \ket{\chi} \otimes \sum_{j=-a}^{a} g(j)\ket{j},
\end{equation}
where the amplitudes are given by a discrete Gaussian profile centered at the origin,
\begin{equation}
g(j) = \mathcal{N}\,\exp\!\left(-\frac{j^2}{4\sigma_0^2}\right),
\end{equation}
where $\sigma_0$ controls the initial width of the Gaussian state and, in the continuum-like regime considered here, coincides with its initial standard deviation to excellent accuracy, while $\mathcal{N}$ is a normalization constant satisfying
\begin{equation}
\sum_{j=-a}^{a} |g(j)|^2 = 1.
\end{equation}
We restrict $\sigma_0$ to values for which the discrete Gaussian is well approximated by its continuum counterpart. Moreover, the bulk size $a$ is chosen sufficiently large that the Gaussian tails are negligible at the boundaries, so that finite-size truncation has no appreciable effect on the shape or effective initial width of the state.

In addition to the localized and Gaussian initial conditions, we consider a truncated-uniform state, in which the walker is initially distributed with equal amplitude over all $N$ sites of the bulk region, i.e.,
\begin{equation}
\ket{\Psi(0)}_U=\ket{\chi}\otimes\sum_{j=-a}^{a}\frac{1}{\sqrt{N}}\ket{j},
\end{equation}
whose initial standard deviation is
\begin{equation}
\sigma_U(0)=\sqrt{\frac{N^2-1}{12}}.
\end{equation}

The truncated-uniform state should be distinguished from a Gaussian state with a large initial width. In the Gaussian case, the probability amplitudes remain spatially weighted according to the Gaussian envelope, whereas the truncated-uniform state assigns exactly the same amplitude to every site within the bulk interval $[-a,a]$ and vanishes outside it. The latter therefore provides a useful limiting case for investigating the dynamics of an initially extended state whose support coincides with the entire bulk region.

Throughout the numerical analysis, we choose $\sigma_0=3$, for which the evolving position distribution no longer exhibits the pronounced multipeak structure characteristic of localized or narrower initial states~\cite{Engster2024high}. This choice yields a sufficiently smooth Gaussian wave-packet dynamics while keeping the initial state well localized relative to the confinement sizes considered, thereby providing an intermediate spatial profile between the localized and truncated-uniform cases. In the finite-size analysis presented below, we consider $N\geq101$. For this $\sigma_0$ value, the Gaussian tails are therefore negligible at the boundaries even for the smallest confinement size considered, so that truncation effects on the initial state can be safely neglected.

These initial conditions provide a useful framework for analyzing how the spatial structure of the walker influences the interplay between delocalization, interference, confinement, and long-time spreading dynamics. All three states are initially separable between the coin and position degrees of freedom, so that $S_E(0)=0$; therefore, any nonzero coin-position entanglement observed during the subsequent evolution is dynamically generated by the quantum walk.

\section{Results and Discussion} \label{secIV}

In the following sections, we characterize the dynamical behavior of the confined quantum walk and introduce the quantities used to analyze its finite-size spreading properties. We first discuss the oscillatory dynamics of the standard deviation, which reflects the repeated reflection and interference of the ballistic components at the coin-flip operations at the boundaries. We then define the spreading-based stiffness in terms of the instantaneous magnitude of the discrete spreading rate. Finally, we introduce the normalized Kullback--Leibler divergence from the uniform distribution as a complementary information-theoretic diagnostic of the long-time redistribution of probability over the physical bulk region.

\subsection{Long-time coherent dynamical behavior}

In a confined quantum walk, the width of the probability distribution exhibits a quasi-periodic oscillatory dynamics generated by the repeated reflection of the ballistic fronts at the coin-flip operations at the boundaries. Since the finite unitary dynamics is not strictly periodic in general, we define an initial effective period rather than an exact recurrence time. A natural estimate for this time scale follows from the maximum group velocity of the Hadamard walk,
\(v_{\max}=1/\sqrt{2}\), which sets the fastest propagation speed of the dominant ballistic components \cite{Kempf2009group}. For a bulk region \(j\in[-a,a]\), these fronts travel a distance of order \(2a\) before returning to the central region. This gives the lower-bound time scale \(T_0=2a/v_{\max}=2\sqrt{2}\,a\). 

\begin{figure*}[htpb]
\centering
\includegraphics[width=\textwidth]{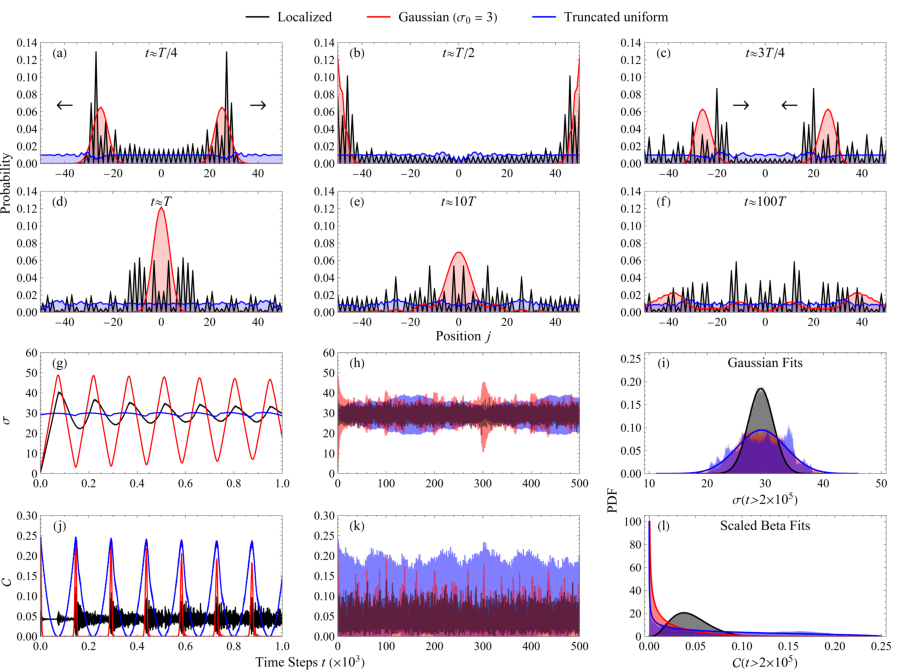}
\caption{
Confinement dynamics and long-time oscillations of a confined discrete-time quantum walk with $N=101$ bulk sites for three different initial spatial states: localized (black), Gaussian with $\sigma_0=3$ (red), and truncated-uniform (blue). (a)--(c) Probability distributions at $t\simeq T/4$, $T/2$, and $3T/4$, respectively, showing the propagation and reflection of the probability fronts at the coin-flip operations at the boundaries; the arrows indicate the dominant direction of motion. (d)--(f) Probability distributions at $t\simeq T$, $10T$, and $100T$, illustrating the progressive development of the confined interference pattern. (g) Early-time evolution of the standard deviation $\sigma(t)$ for the three initial states, revealing pronounced confinement-induced oscillations. (h) Long-time evolution of $\sigma(t)$ up to $5\times10^5$ time steps, where the dynamics remains oscillatory around a well-defined average value. (i) Empirical probability density functions of $\sigma(t)$ evaluated over the interval $2\times10^5<t\leq5\times10^5$, together with Gaussian fits. Despite the different initial spatial profiles, the Gaussian and truncated-uniform states display nearly indistinguishable coarse-grained distributions in this long-time window. (j) Early-time evolution of the coin coherence $\mathcal{C}(t)=|\rho_{LR}(t)|^2$. (k) Corresponding long-time coherence dynamics up to $5\times10^5$ time steps. (l) Empirical probability density functions of $\mathcal{C}(t)$ over the same long-time interval, together with scaled Beta fits, consistent with the bounded support $0\leq \mathcal{C}\leq1/4$.}
\label{Fig2}
\end{figure*}

The actual time scale associated with the first confinement-induced response is generally larger than the ballistic estimate \(T_0\), since the probability distribution contains components with different velocities and undergoes coherent interference upon reflection. Because the detailed evolution depends on the initial spatial state, we define a state-dependent characteristic time \(T\) as the time at which \(\sigma(t)\) reaches its first local minimum. The value of \(T\) is therefore determined independently for the localized, Gaussian, and truncated-uniform initial states and is, in general, different for each case. It should not be interpreted as an exact recurrence period, but rather as an operational time scale characterizing the first dominant confinement-induced variation of the spatial width. In Fig.~\ref{Fig2}, the snapshots labeled \(T/4\), \(T/2\), \(3T/4\), and \(T\) are consequently evaluated separately for each initial state using its corresponding value of \(T\). This state-dependent time rescaling allows the three dynamics to be compared at analogous stages of their early confinement-induced evolution, rather than at identical absolute times.

Figure~\ref{Fig2}(a)--(d) show snapshots of the position distribution during approximately the period \(T\) evaluated separately for each initial state.  Starting from the localized initial state, the walker first develops the two counterpropagating fronts characteristic of ballistic quantum-walk propagation. After reaching the boundaries, the reflected components interfere and the probability profile is driven back toward the central region, completing one expansion--contraction cycle. However, at longer times, as shown in Figs.~\ref{Fig2}(e) and~\ref{Fig2}(f), the instantaneous distribution becomes increasingly irregular due to the accumulation of many interfering reflected components. Nevertheless, the probability remains broadly distributed across the bulk, while the instantaneous occupation continues to reflect the bipartite structure of the walk, with only one parity sublattice populated at each time step. Thus, the confined walk does not converge to a stationary distribution in the classical sense. Instead, it enters a long-time oscillatory regime in which the instantaneous distribution continues to oscillate around the uniform profile, alternating between more localized and more delocalized spatial configurations.

Unlike the localized initial condition, for which the walker alternates between even and odd accessible sites at successive time steps, both delocalized initial states have nonzero amplitudes on the two parity sublattices from the outset. Consequently, both parity sublattices of the bulk remain kinematically accessible throughout the evolution. Their short-time dynamics, however, are markedly different. The Gaussian state is initially concentrated around the center of the lattice and evolves into two smooth counterpropagating wave packets that approximately preserve a Gaussian envelope as they move toward the boundaries. The truncated-uniform state, by contrast, already occupies the entire bulk region with equal amplitude and therefore does not undergo the same expansion process. Since its bulk profile is initially spatially homogeneous, the most prominent dynamical structures are generated by the boundaries, which break translational homogeneity and launch propagating disturbances into the otherwise nearly uniform background. After successive reflections, both delocalized states develop increasingly complex interference patterns, although they retain clear signatures of their distinct initial spatial profiles at early times. Interestingly, these pronounced differences in the transient dynamics become much less evident at long times, where the coarse-grained distributions of the spreading width $\sigma(t)$ for the Gaussian and truncated-uniform states become nearly indistinguishable.

This behavior is also reflected in the dynamics of the standard deviation $\sigma(t)$. Figure~\ref{Fig2}(g) compares its early-time evolution for the localized, Gaussian, and truncated-uniform initial states. For the localized state, $\sigma(t)$ displays the characteristic expansion--contraction dynamics associated with the ballistic propagation of the two probability fronts, their reflection at the boundaries, and their subsequent return toward the central region. The resulting oscillations are initially pronounced but progressively lose their regular large-amplitude structure as repeated reflections generate increasingly complex interference patterns. The Gaussian initial state exhibits an even more pronounced and smoother breathing-like motion: the initially localized Gaussian profile separates into two counterpropagating packets, producing large oscillations of $\sigma(t)$ as the distribution alternates between spatially extended and contracted configurations. In contrast, the truncated-uniform state already occupies the entire bulk region at $t=0$ and therefore does not undergo a comparable global expansion--contraction cycle. Its standard deviation remains close to its initial spatial scale and displays only relatively small oscillations, generated primarily by the boundary-induced propagating disturbances and their subsequent interference. Accordingly, the instantaneous spreading rate $|\sigma(t)-\sigma(t-1)|$ is strongly dependent on the initial spatial state during the transient dynamics: it is largest during the pronounced breathing motion of the Gaussian state, decreases as the localized walk loses its initial ballistic expansion--contraction structure, and remains comparatively suppressed for the truncated-uniform state. Despite these markedly different early-time behaviors, the three cases evolve toward long-time oscillatory regimes centered around similar characteristic spatial widths, as shown in Fig.~\ref{Fig2}(h).

The long-time behavior of $\sigma(t)$ is shown in Fig.~\ref{Fig2}(h). After the initial transient, the standard deviation remains bounded for all three initial states and oscillates around a well-defined characteristic value. These oscillations are substantially smaller than the large early-time excursions, particularly for the localized and Gaussian cases, indicating that the walk has entered a long-time confined regime in which the large-scale expansion--contraction dynamics is strongly reduced. The instantaneous probability distribution nevertheless remains highly time dependent, so the confined walk does not approach a static spatial profile. Instead, its second moment continues to oscillate around a characteristic spatial scale close to that associated with an approximately uniform occupation of the bulk region. Despite the markedly different transient dynamics of the localized, Gaussian, and truncated-uniform initial states, their long-time values of $\sigma(t)$ remain centered around similar spatial widths, while retaining state-dependent oscillatory structures and an apparent slow modulation. The Gaussian and scaled-Beta fits are used here as phenomenological descriptions of the coarse-grained distributions rather than as evidence for an underlying universal probability law.

To characterize the long-time dynamics, an analysis window of $3\times10^5$ time steps was adopted beyond the early-time transient region. This choice is motivated by the presence of an apparent slow modulation in the confined dynamics. When shorter time intervals are considered, the empirical distributions of $\sigma(t)$ retain pronounced structures associated with this modulation, particularly for the Gaussian and truncated-uniform initial states. In contrast, a window of $3\times10^5$ time steps provides a more complete sampling of the long-time dynamics and yields stable Gaussian envelopes for the distributions of $\sigma(t)$. Remarkably, within such long windows, the Gaussian fits obtained for the Gaussian and truncated-uniform initial states become nearly indistinguishable, despite the markedly different spatial profiles of the corresponding initial states, as shown in Fig.~\ref{Fig2}(i). This behavior is robust under translations of the $3\times10^5$-step analysis window, with its initial point varied from $5\times10^4$ to $2\times10^5$. Over these windows, the fitted means for the Gaussian and truncated-uniform cases differ by less than approximately $0.03\%$, while their fitted standard deviations agree within approximately $1.2\%$. The fit parameters for each initial-state case are also nearly insensitive to the position of the analysis window, with variations in the Gaussian widths below approximately $1.5\%$ and variations in the Beta-distribution parameters describing the coin coherence of approximately $1.2\%$ or less. For definiteness, the probability density functions (PDFs) shown in Figs.~\ref{Fig2}(i) and (l) are evaluated over the interval $2\times10^5<t\leq5\times10^5$. These results indicate that averaging over a sufficiently long time window reduces the influence of the apparent slow modulation while revealing a common coarse-grained statistical behavior of the spreading oscillations.

The long-time distribution of the coin coherence is characterized through the quantity $\mathcal{C}(t)=|\rho_{LR}(t)|^2$, where $\rho_{LR}$ is the off-diagonal element of the reduced coin density matrix. Since positivity of the density matrix imposes $|\rho_{LR}|^2\leq\rho_{LL}\rho_{RR}\leq1/4$, the coherence is restricted to the finite interval $0\leq\mathcal{C}\leq1/4$. For this reason, its empirical probability density is fitted by a scaled Beta distribution. Introducing the rescaled variable $x=\mathcal{C}/\mathcal{C}_{\max}=4\mathcal{C}$, with $\mathcal{C}_{\max}=1/4$, the corresponding probability density in terms of the physical coherence variable is
\begin{equation}
p(\mathcal{C})=\frac{4}{B(\alpha,\beta)}(4\mathcal{C})^{\alpha-1}(1-4\mathcal{C})^{\beta-1},
\end{equation}
where $B(\alpha,\beta)$ is the Beta function, which ensures normalization of the distribution. The fitted parameters provide a convenient phenomenological characterization of the shape of the coherence distribution.

For the localized initial state, both fitted shape parameters satisfy $\alpha>1$ and $\beta>1$, yielding a distribution with an interior maximum. In contrast, for the Gaussian and truncated-uniform initial states, the fits yield $\alpha<1$ and $\beta>1$, so that the corresponding Beta densities are strongly weighted toward $\mathcal{C}\rightarrow0$, consistently with the enhanced occurrence of very small coherence values observed numerically. The broader tail of the truncated-uniform case, however, extends to substantially larger values of $\mathcal{C}$, indicating that low-coherence events coexist with comparatively large coherence excursions in the long-time dynamics. For the symmetric initial coin state considered here, the population-imbalance contribution $\mathcal{P}(t)$ remains identically zero throughout the evolution as a consequence of the balanced initial condition, the unbiased Hadamard dynamics in the bulk, and the left--right symmetry preserved by the symmetric reflecting boundaries. Consequently, the long-time internal dynamics is reflected entirely in the coherence contribution. The eigenvalues of the reduced coin state therefore reduce to
\begin{equation}
\lambda_{\pm}(t)=\frac{1}{2}\pm\sqrt{\mathcal{C}(t)},
\end{equation}
so that the coin-position entanglement entropy is completely determined by the coherence contribution $\mathcal{C}(t)$. In particular, the initial product state has $\mathcal{C}(0)=1/4$ and $S_E(0)=0$, whereas the subsequent reduction and temporal variation of $\mathcal{C}(t)$ correspond to dynamically generated coin-position entanglement.

\subsection{Boundary-induced stiffness}

To quantify the resistance of the confined quantum walk to changes in its spatial width, we introduce a spreading-based stiffness measure constructed from the standard deviation $\sigma(t)$. Since the dynamics is discrete in time, the instantaneous magnitude of the spreading rate is naturally characterized by the finite difference
\begin{equation}
\nu_{\sigma}(t)=\left|\sigma(t)-\sigma(t-1)\right|.
\label{eq:vsigma}
\end{equation}
This quantity measures the absolute change in the width of the probability distribution between two consecutive time steps and therefore accounts for both expansion and contraction processes. Large values of $\nu_{\sigma}(t)$ correspond to strong temporal variations of the spatial width, whereas small values indicate suppressed expansion--contraction activity.

A natural reference scale is provided by the corresponding unconfined Hadamard walk, for which the standard deviation grows asymptotically linearly with time,
\begin{equation}
\sigma_{\rm H}(t)\sim \overline{\nu}_{\rm H}\,t,\qquad t\gg1,
\end{equation}
where $\overline{\nu}_{\rm H}$ denotes the asymptotic spreading rate of the free Hadamard walk. This quantity can be obtained analytically by transforming the evolution operator to $k$ space, diagonalizing $U_k$, and evaluating the resulting expressions in the long-time limit \cite{Orthey2019connecting}. For the localized initial state considered here,
\begin{equation}
\overline{\nu}_{\mathrm{H}}^{\,\mathrm{L}}=\sqrt{1-\frac{1}{\sqrt{2}}},
\end{equation}
whereas for delocalized states the asymptotic spreading rate depends on the initial spatial profile. In particular, for Gaussian states its dependence on the initial width $\sigma_0$ follows from the asymptotic expressions derived in Ref.~\cite{Orthey2019connecting}.

Since the spreading stiffness is defined relative to the corresponding free Hadamard walk with the same initial spatial state, the truncated-uniform state requires an $N$-dependent ballistic reference. Its Fourier-space probability density is given by the Fej\'er kernel,
\begin{equation}
\left|\widetilde{f}_{N}(k)\right|^2
=
\frac{1}{N}
\left[
\frac{\sin(Nk/2)}{\sin(k/2)}
\right]^2,
\end{equation}
from which the asymptotic spreading coefficient can be written as
\begin{equation}
\overline{\nu}_{\mathrm{H}}^{\,\mathrm{U}}(N)=\sqrt{I_N},
\label{nuuniform}
\end{equation}
with
\begin{equation}
I_N=1-\frac{1}{\sqrt{2}}-\sqrt{2}\sum_{n=1}^{(N-1)/2}(2\sqrt{2}-3)^n
\left(1-\frac{2n}{N}\right).
\end{equation}
Thus, unlike the localized or fixed-width Gaussian cases, the free reference associated with the truncated-uniform initial condition depends explicitly on $N$, because its spatial support is itself defined by the $N$ sites of the bulk region. In the limit of a very broad uniform state,
\begin{equation}
\lim_{N\rightarrow\infty}\overline{\nu}_{\mathrm{H}}^{\,\mathrm{U}}(N)=
\frac{1}{\sqrt{2}},
\end{equation}
which coincides with the ballistic spreading limit approached by highly delocalized Gaussian states. Numerical values relevant to selected initial states considered here are summarized in Table~\ref{tab.1}.
\begin{table}[b]
\caption{Asymptotic spreading rate of the corresponding free Hadamard walk for the localized, Gaussian ($\sigma_0=3$), and truncated-uniform ($N=101$) states. The coin state is fixed at $(\theta,\varphi)=(\pi/2,\pi/2)$. The localized and Gaussian values are obtained from the asymptotic expressions of Ref.~\cite{Orthey2019connecting}, while the truncated-uniform value follows from the finite-$N$ expression in Eq.~\eqref{nuuniform}.}
\label{tab.1}
\begin{ruledtabular}
\begin{tabular}{lccc}
 & Localized & Gaussian & Truncated-uniform \\
\colrule
$\overline{\nu}_{\rm H}$ & 0.5412 & 0.7028 & 0.7046 \\
\end{tabular}
\end{ruledtabular}
\end{table}

For the localized and fixed-width Gaussian initial states, $\overline{\nu}_{\rm H}$ is independent of the confinement length and therefore provides a fixed ballistic reference scale for comparison with the corresponding confined dynamics. For the truncated-uniform state, by contrast, the reference depends on $N$ because changing the confinement length also changes the support of the initial state. In each case, we use the asymptotic spreading rate of the corresponding free walk and define the dimensionless spreading stiffness as
\begin{equation}
\mathcal{S}_{\nu}(t)
=
\frac{\overline{\nu}_{\rm H}}
{\overline{\nu}_{\rm H}+\nu_{\sigma}(t)}.
\label{eq:stiffness}
\end{equation}
The particular form of $\mathcal{S}_{\nu}(t)$ provides a bounded and dimensionless measure of the instantaneous resistance of the spatial width to temporal variation. Rather than using $\nu_{\sigma}(t)$ directly, the normalization by the asymptotic spreading rate $\overline{\nu}_{\rm H}$ introduces a state-dependent reference scale associated with the corresponding unconfined Hadamard walk. The mapping compresses the normalized width variation into the interval $0<\mathcal{S}_{\nu}\leq1$, with $\mathcal{S}_{\nu}=1/2$ corresponding to an instantaneous width variation equal to the free-walk reference and $\mathcal{S}_{\nu}\rightarrow1$ corresponding to increasingly suppressed width variations. Accordingly, $\mathcal{S}_{\nu}>1/2$ indicates width variations smaller than the corresponding free-walk spreading scale, whereas $\mathcal{S}_{\nu}<1/2$ indicates expansion--contraction activity exceeding this reference. In this sense, $\mathcal{S}_{\nu}$ should be regarded as an operational spreading-based notion of stiffness, inspired by the resistance to boundary-induced redistribution discussed for classical reflecting walks, rather than as a direct quantum analogue of the classical occupation-time stiffness.

\begin{figure}[htpb]
\centering
\includegraphics[width=0.45\textwidth]{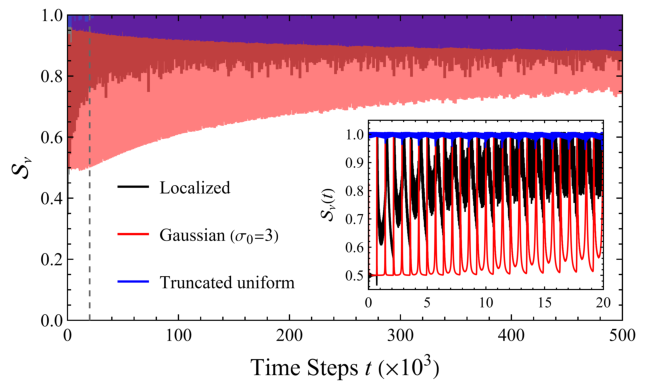}
\caption{Time evolution of the spreading stiffness $\mathcal{S}_{\nu}(t)$ for the localized (black), Gaussian with $\sigma_0=3$ (red), and truncated-uniform (blue) initial states for $N=1001$. The dashed vertical line at $t_c=2\times10^{4}$ marks the end of the transient interval discarded in the long-time averages. The inset shows the initial dynamics over $t\leq t_c$, revealing the coherent oscillatory structure underlying the dense long-time bands observed in the main panel.}
\label{Fig3}
\end{figure}

To identify a suitable time window for the long-time analysis, Fig.~\ref{Fig3} shows the temporal evolution of the spreading stiffness $\mathcal{S}_{\nu}(t)$ for the largest confinement region considered $j\in[-500,500]$. The inset resolves the early-time dynamics and highlights the distinct responses of the three initial states. Up to $t\simeq 5\times10^{2}$, before boundary reflections significantly affect the evolution, the localized and Gaussian initial states remain close to the free-walk reference, with $\mathcal{S}_{\nu}\simeq 1/2$. At later times, boundary reflections and the resulting interference drive repeated excursions above this value, eventually producing the bounded oscillatory dynamics observed in the main panel. By contrast, the truncated-uniform state starts very close to $\mathcal S_\nu=1$ and exhibits bounded oscillations close to unity from the early stages of the evolution. Since this state initially extends over the entire bulk region, it does not display the same free-like ballistic transient observed for the localized and Gaussian states.

The dashed vertical line in Fig.~\ref{Fig3} marks the transient cutoff adopted throughout the finite-size analysis. Since $N=1001$ is the largest confinement size considered, it is also the case in which boundary-induced effects occur on the longest time scale. The observation that the bounded oscillatory regime is already established before $t_c=2\times10^{4}$ therefore provides a conservative criterion for the smaller systems, $101\leq N<1001$, for which the boundaries are reached on shorter time scales. We consequently discard the first $t_c$ time steps for all confinement sizes, ensuring that the subsequent statistics characterize the established bounded oscillatory regime rather than the initial transient. 

Having isolated this long-time regime, we characterize the stiffness through its temporal moments. For each confinement size and initial condition, the stiffness moments are evaluated over the remaining $T_{\rm av}=4.8\times10^{5}$ time steps as
\begin{equation}
\left\langle \mathcal{S}_{\nu}^{m}\right\rangle
=
\frac{1}{T_{\rm av}}
\sum_{t=t_{\rm c}+1}^{t_{\rm c}+T_{\rm av}}
\mathcal{S}_{\nu}^{m}(t),
\end{equation}
where $t_{\rm c}=2\times10^{4}$ is the transient cutoff, and $m=1$ and $m=2$ correspond to the first and second moments, respectively. These moments provide compact measures of the long-time stiffness dynamics and form the basis of the finite-size analysis discussed below.

\begin{figure}[htpb]
\centering
\includegraphics[width=0.45\textwidth]{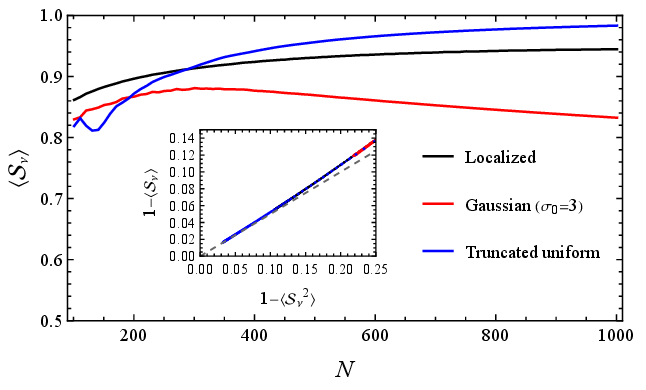}
\caption{Finite-size dependence of the long-time averaged spreading stiffness $\langle\mathcal{S}_{\nu}\rangle$ for the localized (black), Gaussian with $\sigma_0=3$ (red), and truncated-uniform (blue) initial states. The localized and truncated-uniform states exhibit an overall increase of the mean stiffness with the confinement size $N$, whereas the Gaussian state reaches a broad maximum at intermediate sizes and subsequently decreases. The inset shows $1-\langle\mathcal{S}_{\nu}\rangle$ as a function of $1-\langle\mathcal{S}_{\nu}^{2}\rangle$ for the same three initial states. The dashed line represents the high-stiffness reference relation $1-\langle\mathcal{S}_{\nu}\rangle=\frac{1}{2}[1-\langle\mathcal{S}_{\nu}^{2}\rangle]$. For the finite-size analysis, the first $2\times10^{4}$ time steps are discarded and the averages are evaluated over the subsequent $4.8\times10^{5}$ steps.}
\label{Fig4}
\end{figure}

Figure~\ref{Fig4} shows the dependence of the long-time averaged spreading stiffness $\langle\mathcal{S}_{\nu}\rangle$ on the confinement size $N$ for the three initial spatial states, using the temporal averaging procedure defined above. In contrast to the probability-density analysis discussed above, no shorter temporal window is required here, since the objective is to characterize long-time temporal moments rather than the detailed shape of the corresponding distributions.

The three initial states display qualitatively distinct finite-size behaviors. For the localized state, $\langle\mathcal{S}_{\nu}\rangle$ increases overall with $N$ and gradually approaches a high-stiffness regime, indicating that long-time variations of the spatial width become progressively weaker relative to the corresponding ballistic free-walk reference. The Gaussian state exhibits a nonmonotonic dependence: after an initial increase, the mean stiffness reaches a broad maximum at intermediate $N$ and subsequently decreases as the confinement region is enlarged. Since the Gaussian width is kept fixed at $\sigma_0=3$, increasing $N$ moves the reflecting boundaries progressively farther from the initially occupied region, thereby modifying the balance between free spreading and boundary-induced expansion--contraction dynamics. The truncated-uniform state behaves differently, displaying a pronounced finite-size structure at the lower end of the confinement-size range followed by a sustained increase of $\langle\mathcal{S}_{\nu}\rangle$ toward unity. Because this state initially occupies the entire bulk region, the boundary-generated disturbances become progressively less effective in changing its overall spatial width as $N$ increases, resulting in an increasingly stiff long-time response. For the smallest confinement sizes considered, the Gaussian and truncated-uniform states exhibit very similar mean stiffness values, despite their distinct initial spatial profiles. As the confinement size increases, however, their behavior becomes clearly distinguishable: the Gaussian retains the fixed initial width $\sigma_0=3$, whereas the support of the truncated-uniform state increases with $N$. These contrasting trends demonstrate that the boundary-induced stiffness retains a strong dependence on the spatial structure of the initial state even in the long-time regime.

A general consistency relation between the first and second stiffness moments follows by introducing the stiffness deficit $\delta(t)=1-\mathcal{S}_{\nu}(t)$. From $\delta^2=(1-\mathcal{S}_{\nu})^2$ and taking the temporal average, one obtains
\begin{equation}
1-\langle \mathcal{S}_{\nu}^{2}\rangle=2\left(1-\langle \mathcal{S}_{\nu}\rangle\right)-\left\langle\left(1-\mathcal{S}_{\nu}\right)^2\right\rangle .
\end{equation}
In the high-stiffness regime, where $\mathcal{S}_{\nu}(t)$ remains close to unity and
$\langle(1-\mathcal{S}_{\nu})^2\rangle \ll 1-\langle\mathcal{S}_{\nu}\rangle$,
this exact identity reduces to the asymptotic consistency relation
\begin{equation}
1-\langle\mathcal{S}_{\nu}\rangle\simeq\frac{1}{2}\left[1-\langle\mathcal{S}_{\nu}^{2}\rangle\right].
\label{eq:stiffness_high}
\end{equation}
The inset of Fig.~\ref{Fig4} shows this relation for the three initial states. As the dynamics approaches the high-stiffness regime, characterized by $\langle\mathcal{S}_{\nu}\rangle\rightarrow1$, the numerical results tend toward the reference line defined by Eq.~(\ref{eq:stiffness_high}). In this limit, the deficits associated with the first and second stiffness moments, $1-\langle\mathcal{S}_{\nu}\rangle$ and $1-\langle\mathcal{S}_{\nu}^{2}\rangle$, both vanish while satisfying the asymptotic relation above. The agreement with Eq.~(\ref{eq:stiffness_high}) therefore provides a consistency check of the high-stiffness asymptotic regime, rather than a model-specific scaling law.

\subsection{Kullback--Leibler divergence}

A complementary information-theoretic characterization of the confined dynamics can be obtained from the Kullback--Leibler divergence. For the Kullback--Leibler analysis, the two auxiliary reflecting sites are excluded from the spatial distribution used to characterize homogenization. We therefore restrict and renormalize the probability distribution to the $N$ bulk sites, $j\in\mathcal{B}$, before evaluating $\mathcal{D}_{\rm KL}$. Throughout this subsection, $P(j,t)$ denotes this probability distribution restricted and renormalized to the bulk region $\mathcal{B}$. This choice ensures that the Kullback--Leibler divergence quantifies spatial homogenization within the physical bulk region, while the auxiliary sites serve exclusively to implement the unitary reflection.

For a discrete-time quantum walk initialized at a single lattice site, however, the instantaneous probability distribution is restricted by the bipartite structure of the lattice. In particular, the walker occupies only positions having the same parity as the time step, i.e., odd (even) positions $j$ at odd (even) time steps $t$. Consequently, an instantaneous comparison with a uniform distribution over the entire bulk region would artificially assign probability to sites that are kinematically inaccessible at that time.

To remove this parity dependence while preserving the fully unitary dynamics, we introduce a two-step probability distribution constructed from two consecutive time steps. We first define the instantaneous accessible set as
\begin{equation}
\mathcal{A}_t=
\left\{
j\in\mathcal{B}:
|j|\leq\min(t,a),\;
j\equiv t\;(\mathrm{mod}\;2)
\right\},
\end{equation}
and denote by $M(t)=|\mathcal{A}_t|$ the number of accessible sites at time $t$. Since consecutive time steps populate opposite parity sectors, $\mathcal{A}_{t-1}$ and $\mathcal{A}_t$ are disjoint. We therefore define the two-step accessible region as
\begin{equation}
\mathcal{A}_t^{(2)}=\mathcal{A}_{t-1}\cup\mathcal{A}_t,
\end{equation}
whose number of accessible sites is
\begin{equation}
M_2(t)=\left|\mathcal{A}_t^{(2)}\right|=M(t-1)+M(t).
\end{equation}
For the localized initial condition and the nearest-neighbor dynamics considered here, the union of two consecutive parity sectors contains all lattice sites inside the corresponding ballistic light cone, so that
\begin{equation}
\mathcal{A}_t^{(2)}=\left\{j\in\mathcal{B}:|j|\leq\min(t,a)\right\},
\end{equation}
and consequently
\begin{equation}
M_2(t)=2\min(t,a)+1.
\end{equation}
Once the ballistic light cone reaches the boundaries, i.e., for $t\geq a$, the two-step accessible region coincides with the entire bulk, $\mathcal{A}_t^{(2)}=\mathcal{B}$, and hence $M_2(t)=N$.

We define the normalized two-step probability distribution for $t\geq1$ as
\begin{equation}
P^{(2)}(j,t)=\frac{M(t-1)P(j,t-1)+M(t)P(j,t)}{M(t-1)+M(t)}.
\label{eq:Ptwo}
\end{equation}
The weighting by the number of accessible sites ensures that, if the probability is uniformly distributed within each instantaneous parity sector, the resulting two-step distribution is uniform over the combined accessible region $\mathcal{A}_t^{(2)}$. The corresponding uniform reference distribution is therefore
\begin{equation}
P_{\rm unif}^{(2)}(j,t)=
\begin{cases}
1/M_2(t), & j\in\mathcal{A}_t^{(2)},\\[1mm]
0, & j\notin\mathcal{A}_t^{(2)}.
\end{cases}
\end{equation}
The two-step Kullback--Leibler divergence is then
\begin{equation}
D_{\rm KL}^{(2)}(t)=D_{\rm KL}\left[P^{(2)}(t)\|P_{\rm unif}^{(2)}(t)\right],
\end{equation}
or equivalently,
\begin{equation}
D_{\rm KL}^{(2)}(t)=\ln M_2(t)-H_{\mathcal{A}_t^{(2)}}\left[P^{(2)}(t)\right],
\end{equation}
where
\begin{equation}
H_{\mathcal{A}_t^{(2)}}\left[P^{(2)}(t)\right]=-\sum_{j\in\mathcal{A}_t^{(2)}}P^{(2)}(j,t)\ln P^{(2)}(j,t)
\end{equation}
is the Shannon entropy of the two-step probability distribution over the combined accessible region.

For comparison between different confinement sizes, we define the normalized two-step Kullback--Leibler divergence as
\begin{equation}
\mathcal{D}_{\rm KL}(t)=\frac{D_{\rm KL}^{(2)}(t)}{\ln M_2(t)}=1-\frac{H_{\mathcal{A}_t^{(2)}}[P^{(2)}(t)]}{\ln M_2(t)}.
\label{eq:KLtwo}
\end{equation}
The quantity $\mathcal{D}_{\rm KL}(t)$ satisfies $0\leq\mathcal{D}_{\rm KL}(t)\leq1$. For the balanced localized initial condition considered here, $\mathcal{D}_{\rm KL}(1)=0$, since the corresponding two-step distribution is uniform over $\mathcal{A}_1^{(2)}$. Larger values indicate stronger nonuniformity, whereas smaller values indicate a more homogeneous redistribution of probability over the corresponding accessible bulk region.

The parity correction is not required for the Gaussian and truncated-uniform initial states. Unlike the localized state, these delocalized states have initial support on both parity sublattices and therefore do not alternate between mutually exclusive even and odd accessible sectors. The spatial reference region used in the Kullback--Leibler analysis can thus be taken as the complete bulk $\mathcal{B}$, containing $N$ sites, at every time step. For these states, the Kullback--Leibler divergence is evaluated directly from the instantaneous probability distribution $P(j,t)$ relative to the uniform distribution $1/N$, yielding
\begin{equation}
D_{\rm KL}(t)=\ln N-H[P(t)],
\end{equation}
where
\begin{equation}
H[P(t)]=-\sum_{j=-a}^{a}P(j,t)\ln P(j,t).
\end{equation}
Accordingly, the normalized divergence used for the delocalized initial states is
\begin{equation}
\mathcal{D}_{\rm KL}(t)=1-\frac{H[P(t)]}{\ln N}.
\label{eq:KLdelocalized}
\end{equation}
Thus, the localized and delocalized cases employ the same normalized measure of departure from a uniform distribution, but with the reference support adapted to their different kinematic accessibility: a two-step combined parity distribution is required for the localized state, whereas the instantaneous distribution can be used directly for the Gaussian and truncated-uniform states.

The normalized Kullback--Leibler divergence is used here as a complementary redistribution diagnostic. It quantifies the spatial nonuniformity of the probability distribution over the corresponding bulk reference region, whereas the spreading stiffness $\mathcal{S}_{\nu}$ characterizes the resistance of the spatial width to temporal variation. The two quantities therefore probe distinct aspects of the confined quantum dynamics and exhibit distinct dependences on the confinement length.

\begin{figure}[htpb]
\centering
\includegraphics[width=0.45\textwidth]{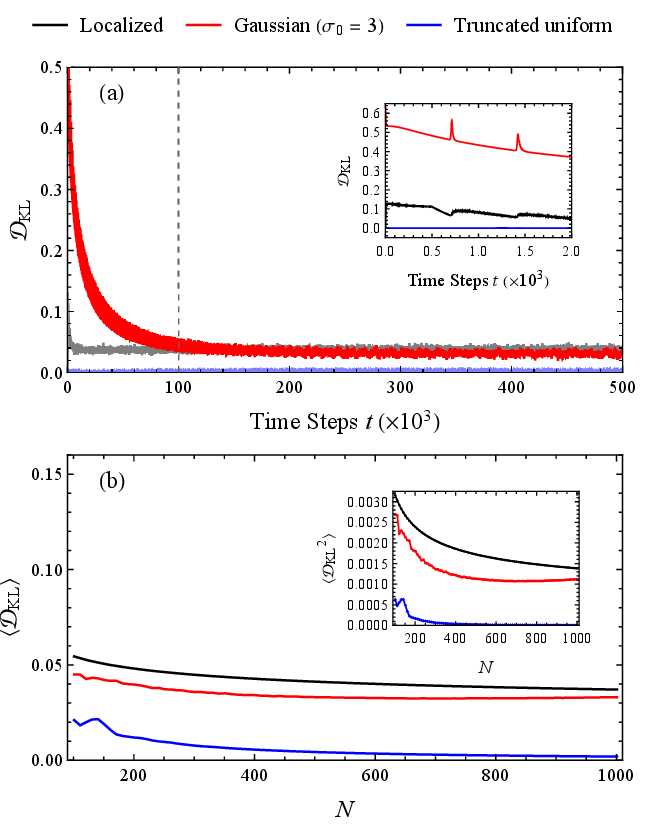}
\caption{Kullback--Leibler analysis for the localized (black), Gaussian with $\sigma_0=3$ (red), and truncated-uniform (blue) initial states. (a) Time evolution of the normalized Kullback--Leibler divergence $\mathcal{D}_{\rm KL}(t)$ for $N=1001$. The dashed vertical line at $t=10^{5}$ marks the transient cutoff adopted for the long-time Kullback--Leibler statistics. The inset resolves the early-time dynamics. (b) Finite-size dependence of the temporal mean $\langle\mathcal{D}_{\rm KL}\rangle$, evaluated after discarding the first $10^{5}$ time steps. The inset shows the corresponding second temporal moment $\langle\mathcal{D}_{\rm KL}^{2}\rangle$ as a function of $N$.}
\label{Fig5}
\end{figure}

Figure~\ref{Fig5} provides an additional characterization of the confined dynamics through the normalized Kullback--Leibler divergence. An important difference with respect to the stiffness analysis is the considerably longer transient time scale. While the bounded oscillatory regime of $\mathcal{S}_{\nu}(t)$ is already established for $t_c=2\times10^{4}$, the systematic decrease of $\mathcal{D}_{\rm KL}(t)$ persists up to approximately $t\simeq10^{5}$, as indicated by the dashed vertical line in Fig.~\ref{Fig5}(a). This separation of time scales shows that the stabilization of the width dynamics precedes the slower spatial reorganization measured by the Kullback--Leibler divergence. We therefore adopt $t_c=10^{5}$ for the long-time Kullback--Leibler statistics.

The inset of Fig.~\ref{Fig5}(a) further shows that the initial decrease of $\mathcal{D}_{\rm KL}(t)$ is not strictly monotonic. In particular, the localized and Gaussian states exhibit boundary-induced nonmonotonic features and sharp excursions when the spreading distribution first interacts with the reflecting boundaries. These features mark the onset of the confinement-induced spatial reorganization before the system reaches its long-time regime. The truncated-uniform state behaves differently: since it initially occupies the entire bulk region, its Kullback--Leibler divergence remains close to zero throughout the evolution. Consequently, $\mathcal{D}_{\rm KL}$ is only weakly sensitive to dynamical changes of this state as long as its spatial distribution remains close to uniform, illustrating that spatial homogenization and dynamical stiffness probe different aspects of the confined walk.

The finite-size behavior shown in Fig.~5(b) also retains a clear dependence on the initial spatial state. The localized and Gaussian states exhibit comparable mean values of $\mathcal{D}_{\rm KL}$ and similar decreasing trends with increasing $N$, whereas the truncated-uniform state remains markedly lower and approaches values close to zero more rapidly. This distinction reflects their different initial spatial structures: the localized and Gaussian states are initially concentrated within a restricted region of the bulk, while the truncated-uniform state extends over the entire bulk from the outset. The temporal mean and second moment of $\mathcal{D}_{\rm KL}$ display qualitatively similar finite-size dependences, preserving the same relative ordering of the three initial states over the confinement-size range considered. Interestingly, this organization differs from that observed for the spreading stiffness, for which the Gaussian and truncated-uniform states exhibit similar mean values at the lower end of the confinement-size range. The Kullback--Leibler statistics therefore reinforce the distinct character of the two measures: $\mathcal{S}_{\nu}$ characterizes resistance to temporal variations of the spatial width, whereas $\mathcal{D}_{\rm KL}$ quantifies the degree of spatial nonuniformity of the probability distribution.

As a robustness check, we also evaluated the Kullback--Leibler statistics of the Gaussian and truncated-uniform states after applying an analogous two-step temporal coarse graining to that used for the localized state. The resulting finite-size dependences preserve the relative ordering and qualitative trends shown in Fig.~\ref{Fig5}(b). In particular, the temporal standard deviation of $\mathcal{D}_{\rm KL}$ (not shown) changes by less than approximately $0.35\%$ for the Gaussian state and $0.23\%$ for the truncated-uniform state, while the corresponding second moments change by less than approximately $3.2\%$ and $3.6\%$, respectively. These results indicate that the differences observed among the three initial states are not an artifact of the different temporal treatments required by the parity constraint of the localized walk.

\section{Conclusion}\label{secV}

We investigated the long-time dynamics of a discrete-time quantum walk under unitary spatial confinement implemented through Pauli-$X$ coin-flip operations at the boundaries. Within this framework, we introduced a dimensionless spreading stiffness, $\mathcal{S}_{\nu}(t)$, which compares the instantaneous variation of the spatial width with the asymptotic spreading scale of the corresponding unconfined Hadamard walk. The confined dynamics remains intrinsically nonstationary, with persistent coherent oscillations and interference rather than relaxation to a static distribution. Although long temporal sampling reveals some common coarse-grained features of the confined dynamics, particularly in the characteristic spatial widths, the response does not become independent of the initial state. 

The finite-size stiffness analysis shows that the long-time dynamics retains a strong dependence on the initial spatial profile. The localized and truncated-uniform states show an overall increase in stiffness as the confinement size increases, whereas the Gaussian state exhibits a nonmonotonic dependence with a broad maximum at intermediate sizes. At the lower end of the confinement-size range, the Gaussian and truncated-uniform states exhibit similar mean stiffness despite their markedly different initial spatial profiles, while their responses separate as the confinement size increases. The first and second stiffness moments also approach the expected asymptotic consistency relation in the high-stiffness regime, providing a useful check of the suppression of width variations.

The Kullback--Leibler analysis reveals a distinct aspect of the confined dynamics. The spatial redistribution measured by $\mathcal{D}_{\rm KL}$ evolves on a longer time scale than the stiffness, indicating that the establishment of the long-time width-oscillation regime precedes the slower spatial reorganization. The localized and Gaussian states exhibit comparable levels of spatial nonuniformity and similar decreasing trends as the confinement size $N$ increases. By contrast, the truncated-uniform state remains much closer to the uniform reference, with substantially smaller temporal mean and second moment of $\mathcal{D}_{\rm KL}$ that decrease more rapidly with $N$. Its persistently small values of $\mathcal{D}_{\rm KL}$ also make this measure weakly sensitive to dynamical rearrangements that preserve an almost uniform spatial distribution. These contrasting trends further demonstrate that dynamical stiffness and spatial homogenization characterize distinct properties of the confined walk.

The reduced coin dynamics exhibits nontrivial coherence statistics and dynamically generated coin-position entanglement. For the symmetric coin state considered here, the population imbalance remains identically zero, so that the eigenvalues of the reduced coin state, and hence the entanglement entropy, are determined entirely by the coherence contribution. Overall, the results show that boundary-induced stiffness emerges from the interplay between reflecting boundaries, system size, and the spatial structure of the initial state. The stiffness and Kullback--Leibler divergence therefore provide complementary measures of temporal resistance to width variations and spatial nonuniformity, respectively. 

Future work may extend this analysis by introducing measurement protocols, decoherence, or other mechanisms that interpolate between coherent quantum dynamics and classical stochastic behavior. Such extensions could provide a closer connection with occupation-time statistics and the classical L\'evy--arcsine picture of stiffness. It would also be interesting to investigate alternative boundary protocols, different coin operators, and higher-dimensional geometries, and to determine how these modifications affect the relation between dynamical stiffness and spatial homogenization identified here. Such studies may further clarify how these distinct aspects of confined quantum dynamics can be controlled independently.

\section*{Acknowledgments}

R. Vieira thanks Fundação de Amparo à Pesquisa e Inovação do Estado de Santa Catarina (FAPESC), Edital 25/2025 for funding.

\end{document}